\documentclass[prd,superscriptaddress,amsfonts,amssymb,amsmath,showpacs,twocolumn]{revtex4-2}
\usepackage{bm}
\usepackage{amsfonts}
\usepackage{latexsym}
\usepackage[latin1]{inputenc}
\usepackage{graphicx}
\usepackage{amsmath}
\usepackage{palatino}
\usepackage{ragged2e}
\usepackage{mathpazo}
\usepackage{textcomp}
\usepackage{float}
\usepackage{booktabs}
\usepackage{dcolumn}
\usepackage{multirow}
\usepackage{hyperref}
\hypersetup{colorlinks,citecolor=blue}
\usepackage{amsmath}
\usepackage{xcolor}
\usepackage{orcidlink}
\usepackage[caption=false]{subfig}
\usepackage{commath}
\def\jnl@style{\it}
\def\aaref@jnl#1{{\jnl@style#1}}

\def\aaref@jnl#1{{\jnl@style#1}}

\def\aj{\aaref@jnl{AJ}}                   
\def\apj{\aaref@jnl{ApJ}}                 
\def\apjl{\aaref@jnl{ApJ}}                
\def\apjs{\aaref@jnl{ApJS}}               
\def\apss{\aaref@jnl{Ap\&SS}}             
\def\aap{\aaref@jnl{A\&A}}                
\def\aapr{\aaref@jnl{A\&A~Rev.}}          
\def\aaps{\aaref@jnl{A\&AS}}              
\def\mnras{\aaref@jnl{Mon.~Not.~Roy.~Astron.~Soc.}}             
\def\prd{\aaref@jnl{Phys.~Rev.~D}}        
\def\prc{\aaref@jnl{Phys.~Rev.~C}}  
\def\prl{\aaref@jnl{Phys.~Rev.~Lett.}}    
\def\qjras{\aaref@jnl{QJRAS}}             
\def\skytel{\aaref@jnl{S\&T}}             
\def\ssr{\aaref@jnl{Space~Sci.~Rev.}}     
\def\zap{\aaref@jnl{ZAp}}                 
\def\nat{\aaref@jnl{Nature}}              
\def\aplett{\aaref@jnl{Astrophys.~Lett.}} 
\def\apspr{\aaref@jnl{Astrophys.~Space~Phys.~Res.}} 
\def\physrep{\aaref@jnl{Phys.~Rep.}}      
\def\physscr{\aaref@jnl{Phys.~Scr}}       
\def\commat{\aaref@jnl{Comm.~Math.~Phys.}}              
\def\science{\aaref@jnl{Science}}               
\def\cqg{\aaref@jnl{Classical Quant.~Grav.}}            
\def\jpcs{\aaref@jnl{JPCS}}                                     
\def\ijmpd{\aaref@jnl{Int.~J.~Mod.~Phys.~D}}                    
\def\grg{\aaref@jnl{Gen.~Relat.~Gravit.}}               
\def\rpp{\aaref@jnl{Rep.~Prog.~Phys.}}          
\def\npa{\aaref@jnl{Nucl.~Phys.~A}}        
\def\lrr{\aaref@jnl{Living Rev.~Rel.}}                   
\def\jcap{\aaref@jnl{J.~Cosmology Astropart.~Phys.}}    
\def\rmp{\aaref@jnl{Rev.~Mod.~Phys.}}   
\def\epjc{\aaref@jnl{Eur.~Phys.~J.~C}}

\allowdisplaybreaks[1]

\begin{document}

\color{black}       

\title{Cosmology in $f(R,L_m)$ gravity}

\author{Lakhan V. Jaybhaye\orcidlink{0000-0003-1497-276X}}
\email{lakhanjaybhaye@gmail.com}
\affiliation{Department of Mathematics, Birla Institute of Technology and
Science-Pilani,\\ Hyderabad Campus, Hyderabad-500078, India.}
\author{Raja Solanki\orcidlink{0000-0001-8849-7688}}
\email{rajasolanki8268@gmail.com}
\affiliation{Department of Mathematics, Birla Institute of Technology and
Science-Pilani,\\ Hyderabad Campus, Hyderabad-500078, India.}
\author{Sanjay Mandal\orcidlink{0000-0003-2570-2335}}
\email{sanjaymandal960@gmail.com}
\affiliation{Department of Mathematics, Birla Institute of Technology and
Science-Pilani,\\ Hyderabad Campus, Hyderabad-500078, India.}
\author{P.K. Sahoo\orcidlink{0000-0003-2130-8832}}
\email{pksahoo@hyderabad.bits-pilani.ac.in}
\affiliation{Department of Mathematics, Birla Institute of Technology and
Science-Pilani,\\ Hyderabad Campus, Hyderabad-500078, India.}

\date{\today}
\begin{abstract}

In this letter, we investigate the cosmic expansion scenario of the universe in the framework of $f(R,L_m)$ gravity theory.  We consider a non-linear $f(R,L_m)$ model, specifically, $f(R,L_m)=\frac{R}{2}+L_m^n + \beta$, where $n$ and $\beta$ are free model parameters. Then we derive the motion equations for flat FLRW universe and obtain the exact solution of corresponding field equations. Then we estimate the best fit ranges of model parameters by using updated $H(z)$ datasets consisting of 57 points and the Pantheon datasets consisting of 1048 points. Further we investigate the physical behavior of density and the deceleration parameter. The evolution of deceleration parameter depicts a transition from deceleration to acceleration phases of the universe. Moreover, we analyze the stability of the solution of our cosmological model under the observational constraint by considering a linear perturbation. Lastlty, we investigate the behavior of Om diagnostic parameter and we observe that our model shows quintessence type behavior. We conclude that our $f(R,L_m)$ cosmological model agrees with the recent observational studies and can efficiently describe the late time cosmic acceleration.

\end{abstract}

\maketitle

\section{Introduction}\label{sec1}
\justify 

Recent observations of type Ia supernovae \cite{Riess,Perlmutter} together with observational studies of the Sloan Digital Sky Survey \cite{sdss}, Wilkinson Microwave Anisotropy Probe \cite{D.N.}, Baryonic Acoustic Oscillations \cite{D.J.,W.J.}, Large scale Structure \cite{T.Koivisto,S.F.}, and the Cosmic Microwave Background Radiation \cite{C.L.,R.R.} indicates accelerating behavior of expansion phase of the universe. The standard cosmology strongly supported the dark energy models as resolution of this fundamental question. The most prominent description of dark energy is the cosmological constant $\Lambda$ that can be associated to the vacuum quantum energy \cite{S.W.}. Even though cosmological constant $\Lambda$ fits well with observational data, it is suffering with two major issues namely coincidence problem and cosmological constant problem \cite{E.J.}. Its value obtained from Particle Physics has discrepancy of nearly 120 orders of magnitude with its value required to fit the cosmological observations. Another promising way to describe the recent observations on cosmic expansion scenario of the universe is to consider that the Einstein's general relativity models break downs at large cosmic scales and a more generic action characterizes the gravitational field. There are several ways to generalize the Einstein-Hilbert action of general relativity. The theoretical models in which the standard action is replaced by the generic function $f(R)$, where $R$ is Ricci scalar, introduced in \cite{H.A.,R.K.,H.K.}. The description of late time expansion scenario can be achieved by $f(R)$ gravity \cite{Carr} and the constraints of viable cosmological models have been explored in \cite{Cap,LAM}. The viable $f(R)$ gravity models in the context of solar system tests do exist \cite{Noj,V.F.,P.J.,L.A.}. Observational signatures of $f(R)$ dark energy models along with the solar system and equivalence principle constraints on $f(R)$ gravity have been presented in \cite{Shin,Liu,Sal,Sean,Alex}. Another $f(R)$ models that unifies the early inflation with dark energy and passes through local tests have been discussed in \cite{Noj-2,Noj-3,G.C.}. Moreover, one can check the references \cite{JS,SC,RC} for various cosmological implications of $f(R)$ gravity models.

An extension of the $f(R)$ gravity theory that includes an explicit coupling of matter Lagrangian density $L_m$ with generic function $f(R)$ was proposed in \cite{O.B.}. As a consequence of this matter-geometry coupling, an extra force orthogonal to four velocity vector appears with non-geodesic motion of the massive particles. This model was extended to the case of the arbitrary couplings in both matter and geometry \cite{THK}. The cosmological and astrophysical implications of the non-minimal matter-geometry couplings have been extremely investigated in \cite{THK-2,THK-3,SNN,V.F.-2,V.F.-3}. Recently, Harko and Lobo \cite{THK-4} proposed more evolved generalization of matter-curvature coupling theories called $f(R,L_m)$ gravity theory, where $f(R,L_m)$ represents an arbitrary function of the matter Lagrangian density $L_m$ and the Ricci scalar $R$. The $f(R,L_m)$ gravity theory can be considered as the maximal extension of all the gravitational theories constructed in Reimann space. The motion of test particles in $f(R,L_m)$ gravity theory is non-geodesic and an extra force orthogonal to four velocity vector arises. The $f(R,L_m)$ gravity models admits an explicit violation of the equivalence principle, which is higly constrained by solar system tests \cite{FR,JP}. Recently, Wang and Liao have studied energy conditions in $f(R,L_m)$ gravity \cite{WG}. Gonclaves and Moraes analyzed cosmology from non-minimal matter geometry coupling by taking into account the  $f(R,L_m)$ gravity \cite{GM}. 

The present letter is organized as follows. In Sec \ref{sec2}, we present the fundamental formulation of $f(R,L_m)$ gravity. In Sec \ref{sec3}, we derive the motion equations for the flat FLRW universe. In Sec \ref{sec4}, we consider a cosmological $f(R,L_m)$ model and then we derive the expression for Hubble parameter and the deceleration parameter. In the next section Sec \ref{sec5}, we find the best ranges of the model parameters by using $H(z)$, Pantheon, and the combine H(z)+Pantheon data sets. Further, we analyze the behavior of cosmological parameters for the values of model parameters constrained by the observational data sets. Moreover in Sec \ref{sec6}, we investigate stability of obtained solution under the observational constraint by assuming a linear perturbation of the Hubble parameter. Further in sec \ref{sec7}, we employ Om daignostic test to differentiate our cosmological model with other models of dark energy, Finally in Sec \ref{sec8}, we discuss and conclude our results.

\section{ $f(R,L_m)$ Gravity Theory}\label{sec2}

\justify
The following action governs the gravitational interactions in $f(R,L_m)$ gravity,

\begin{equation}\label{1}
S= \int{f(R,L_m)\sqrt{-g}d^4x} 
\end{equation}

where $f(R,L_m)$ represents an arbitrary function of the Ricci scalar $R$ and the matter Lagrangian term $L_m$. 

The Ricci scalar $R$ can be obtained by contracting the Ricci tensor $R_{\mu\nu}$ as

\begin{equation}\label{2}
R= g^{\mu\nu} R_{\mu\nu}
\end{equation} 

where the Ricci tensor is defined by
 
\begin{equation}\label{3}
R_{\mu\nu}= \partial_\lambda \Gamma^\lambda_{\mu\nu} - \partial_\mu \Gamma^\lambda_{\lambda\nu} + \Gamma^\lambda_{\mu\nu} \Gamma^\sigma_{\sigma\lambda} - \Gamma^\lambda_{\nu\sigma} \Gamma^\sigma_{\mu\lambda}
\end{equation}

Here $\Gamma^\alpha_{\beta\gamma}$ represents the components of the well-known Levi-Civita connection defined by

\begin{equation}\label{4}
\Gamma^\alpha_{\beta\gamma}= \frac{1}{2} g^{\alpha\lambda} \left( \frac{\partial g_{\gamma\lambda}}{\partial x^\beta} + \frac{\partial g_{\lambda\beta}}{\partial x^\gamma} - \frac{\partial g_{\beta\gamma}}{\partial x^\lambda} \right)
\end{equation}

Now one can acquired the following field equation by varying the action \eqref{1} for the metric tensor $g_{\mu\nu}$,

\begin{equation}\label{5}
f_R R_{\mu\nu} + (g_{\mu\nu} \square - \nabla_\mu \nabla_\nu)f_R - \frac{1}{2} (f-f_{L_m}L_m)g_{\mu\nu} = \frac{1}{2} f_{L_m} T_{\mu\nu}
\end{equation}

Here $f_R \equiv \frac{\partial f}{\partial R}$, $f_{L_m} \equiv \frac{\partial f}{\partial L_m}$, and $T_{\mu\nu}$ represents the energy-momentum tensor for the perfect type fluid, defined by 

\begin{equation}\label{6}
T_{\mu\nu} = \frac{-2}{\sqrt{-g}} \frac{\delta(\sqrt{-g}L_m)}{\delta g^{\mu\nu}}
\end{equation}

The relation between the trace of energy-momentum tensor $T$, Ricci scalar $R$, and the Lagrangian density of matter $L_m$ obtained by contracting the field equation \eqref{5}

\begin{equation}\label{7}
R f_R + 3\square f_R - 2(f-f_{L_m}L_m) = \frac{1}{2} f_{L_m} T
\end{equation}

Here $\square F = \frac{1}{\sqrt{-g}} \partial_\alpha (\sqrt{-g} g^{\alpha\beta} \partial_\beta F)$ for any scalar function $F$ .

Moreover, one can acquired the following result by taking covariant derivative in equation \eqref{5}

\begin{equation}\label{8}
\nabla^\mu T_{\mu\nu} = 2\nabla^\mu ln(f_{L_m}) \frac{\partial L_m}{\partial g^{\mu\nu}}
\end{equation} \\

\section{Motion equations in $f(R,L_m)$ gravity}\label{sec3}

Taking into account the spatial isotropy and homogeneity of our universe, we assume the following flat FLRW metric \cite{Ryden} for our analysis

\begin{equation}\label{9}
ds^2= -dt^2 + a^2(t)[dx^2+dy^2+dz^2]
\end{equation}

Here, $ a(t) $ is the scale factor that measures the cosmic expansion at a time $t$. For the line element \eqref{9}, the non-vanishing components of Christoffel symbols are

\begin{equation}\label{10}
\Gamma^0_{ij}= a\dot{a}\delta_{ij} , \: \:  \Gamma^k_{0j}= \Gamma^k_{j0}= \frac{\dot{a}}{a} \delta^k_j
\end{equation}
here $i,j,k=1,2,3$. 

Using equation \eqref{3}, we get the non-zero components of Ricci tensor as

\begin{equation}\label{11}
R_{00}=-3 \frac{\ddot{a}}{a} \: , \: R_{11}=R_{22}=R_{33}= a \ddot{a}+2 \dot{a}^2
\end{equation}

Hence the Ricci scalar obtained corresponding to the line element \eqref{9} is

\begin{equation}\label{12}
R= 6 \frac{\ddot{a}}{a}+ 6 \bigl( \frac{\dot{a}}{a} \bigr)^2 = 6 ( \dot{H}+2H^2 )
\end{equation}
Here $H=\frac{\dot{a}}{a}$ is the Hubble parameter.

The energy-momentum tensor characterizing the universe filled with perfect fluid type matter-content for the line element \eqref{9} is given by,

\begin{equation}\label{13}
\mathcal{T}_{\mu\nu}=(\rho+p)u_\mu u_\nu + pg_{\mu\nu}
\end{equation}

Here $\rho$ is the matter-energy density, $p$ is the spatially isotropic pressure, and $u^\mu=(1,0,0,0)$ are components of the four velocities of the cosmic perfect fluid.

The Friedmann equations that describes the dynamics of the universe in $f(R,L_m)$ gravity reads as

\begin{equation}\label{14}
3H^2 f_R + \frac{1}{2} \left( f-f_R R-f_{L_m}L_m \right) + 3H \dot{f_R}= \frac{1}{2}f_{L_m} \rho 
\end{equation}
and
\begin{equation}\label{15}
\dot{H}f_R + 3H^2 f_R - \ddot{f_R} -3H\dot{f_R} + \frac{1}{2} \left( f_{L_m}L_m - f \right) = \frac{1}{2} f_{L_m}p
\end{equation}

\section{Cosmological $f(R,L_m)$ Model }\label{sec4}

We consider the following functional form \cite{LB} for our analysis,

\begin{equation}\label{16} 
f(R,L_m)=\frac{R}{2}+L_m^n + \beta 
\end{equation}

where $\beta$ and $n$ are free model parameters. \\

Then for this particular $f(R,L_m)$ model with $L_m=\rho$ \cite{HLR}, the Friedmann  equations \eqref{14} and \eqref{15} for the matter dominated universe becomes

\begin{equation}\label{17}
3H^2=(2n-1) \rho^n-\beta
\end{equation}

and

\begin{equation}\label{18}
2\dot{H}+3H^2+\beta=(n-1)\rho^n
\end{equation}

Further, one can acquire the following matter conservation equation by taking trace of the field equations

\begin{equation}\label{c}
n\dot{\rho}+ 3H \rho = 0
\end{equation}

In particular, for $n=1$ and $\beta=0$ one can retrieve the usual Friedmann equations of GR.\\

From equation \eqref{17} and \eqref{18}, we have

\begin{equation}\label{19}
\dot{H}+\frac{3n}{2(2n-1)} H^2+\frac{n}{2(2n-1)}\beta=0
\end{equation}

Then by using $ \frac{1}{H} \frac{d}{dt}= \frac{d}{dln(a)}$, we have the following first-order differential equation

\begin{equation}\label{20}
\frac{dH}{dln(a)}+ \frac{3n}{2(2n-1)}H =-\frac{n\beta}{2(2n-1)}\frac{1}{H}
\end{equation}

Now by integrating the above equation, one can obtained the expression for Hubble parameter in terms of redshift as follows

\begin{equation}\label{21}
H(z)=\bigl[ H_0^2 (1+z)^{\frac{3n}{2n-1}}+\frac{\beta}{3} \{ (1+z)^{\frac{3n}{2n-1}}-1 \} \bigr]^\frac{1}{2}
\end{equation}

Here $H_0$ is the present value of the Hubble parameter.

The deceleration parameter plays a vital role to describe the dynamics of expansion phase of the universe and it is defined as

\begin{equation}\label{22}
q(z)=-1-\frac{\dot{H}}{H^2}
\end{equation}

By using \eqref{21} in \eqref{22}, we have

\begin{equation}\label{23}
q(z)=-1+\frac{3n(3H_0^2+\beta)}{2(2n-1) \{ 3H_0^2+\beta [ 1-(1+z)^{\frac{3n}{1-2n}} ] \} }
\end{equation}

\section{Observational Constraints}\label{sec5}

In this section, we analyze the observational aspects of our cosmological model. We use the $H(z)$ data sets and Pantheon data sets to find the best fit ranges of the model parameters $n$ and $\beta$. To constrain the model parameters, we employ the standard Bayesian technique and likelihood function along with the Markov Chain Monte Carlo (MCMC) method in \texttt{emcee} python library \cite{Mackey/2013}. We use the following probability function to maximize the best fit ranges of the parameters

\begin{equation}\label{24}
\mathcal{L} \propto exp(-\chi^2/2)
\end{equation} 

Here $\chi^2$ represents pseudo chi-sqaured function \cite{BS}. The $\chi^2$ function used for different data sets are given below.

\subsubsection{H(z) datasets}

In this work, we have taken an updated set of $57$ data points of $H(z)$ measurements in the range of redshift given as $0.07 \leq z \leq 2.41$ \cite{GSS}. In general, there are two well established techniques to measure the values of $H(z)$ at given redshift namely line of sight BAO \cite{H1,H2,H3,H4,H5} and the differential age technique \cite{D1,D2,D3,D4}. For the complete list of data points, see the reference \cite{RS}. Moreover, we have taken $H_0 = 69$ Km/s/Mpc for our analysis \cite{planck_collaboration/2020}. To estimate the mean values of the model parameters $n$ and $\beta$, we define the chi-square function as follows,

\begin{equation}\label{25}
\chi _{H}^{2}(n,\beta)=\sum\limits_{k=1}^{57}
\frac{[H_{th}(z_{k},n,\beta)-H_{obs}(z_{k})]^{2}}{
\sigma _{H(z_{k})}^{2}}.  
\end{equation}

Here, $H_{th}$ denotes the theoretical value of the Hubble parameter obtained by our model whereas $H_{obs}$ represents its observed value and $\sigma_{H(z_{k})}$ represents the standard deviation. The $1-\sigma$ and $2-\sigma$ likelihood contours for the model parameters $n$ and $\beta$ using $H(z)$ data sets is presented below.

\begin{figure}[H]
\includegraphics[scale=0.85]{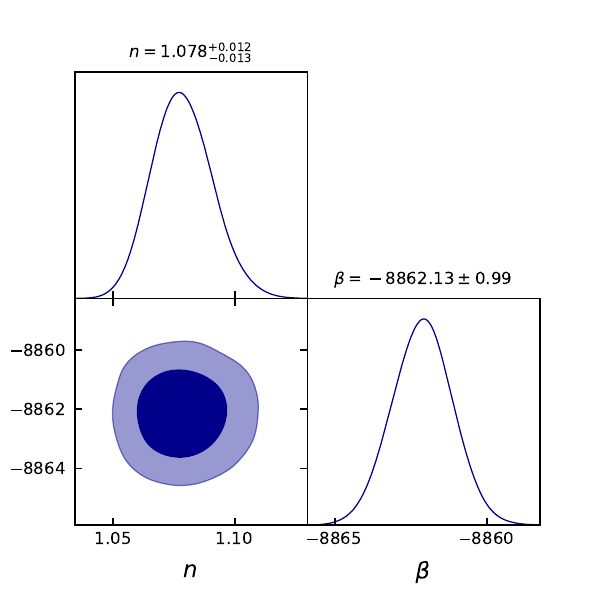}
\caption{The $1-\sigma$ and $2-\sigma$ likelihood contours for the model parameters using $H(z)$ data sets}\label{f1}
\end{figure}

The obtained best fit ranges of the model parameters are $n=1.078^{+0.012}_{-0.013}$ and $\beta=-8862.13 \pm 0.99$.

\subsubsection{Pantheon datasets}

Recently, Pantheon supernovae type Ia data samples consisting of 1048 data points have been released. The PanSTARSS1 Medium, SDSS, SNLS, Deep Survey, numerous low redshift surveys and HST surveys contribute to it. Scolnic et al. \cite{Scolnic/2018} put together the Pantheon supernovae type Ia samples consisting of 1048 in the redshift range $z \in [0.01,2.3]$.  For a spatially  flat universe \cite{planck_collaboration/2020}, the luminosity distance reads as

\begin{equation}\label{26}
D_{L}(z)= (1+z) \int_{0}^{z} \frac{c dz'}{H(z')},
\end{equation}
Here $c$ is the speed of light.

For statistical analysis, the $\chi^{2}$ function for supernovae samples is obtained by correlating the theoretical distance modulus 

\begin{equation}\label{27}
\mu(z)= 5log_{10}D_{L}(z)+\mu_{0}, 
\end{equation}
with 
\begin{equation}\label{28}
\mu_{0} =  5log(1/H_{0}Mpc) + 25,
\end{equation}
such that
\begin{equation}\label{29}
\chi^2_{SN}(p_1,....)=\sum_{i,j=1}^{1048}\bigtriangledown\mu_{i}\left(C^{-1}_{SN}\right)_{ij}\bigtriangledown\mu_{j},
\end{equation}

Here $p_j$ denotes the free model parameters and $C_{SN}$ represents the covariance metric \cite{Scolnic/2018}, and
 \begin{align*}
  \quad \bigtriangledown\mu_{i}=\mu^{th}(z_i,p_1,...)-\mu_i^{obs}.
 \end{align*}
 
where $\mu_{th}$ is theoretical value of the distance modulus whereas $\mu_{obs}$  its observed value.

We have obtained the best fit ranges for parameters $n$ and $\beta$ of our model by minimizing the chi-square function for the supernovae samples. The $1-\sigma$ and $2-\sigma$ likelihood contours for the model parameters $n$ and $\beta$ using Pantheon data sample is presented below. 

\begin{figure}[H]
\includegraphics[scale=0.85]{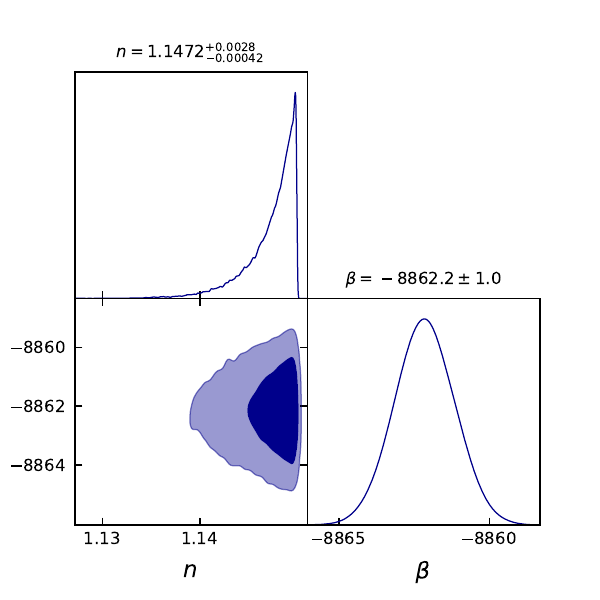}
\caption{The $1-\sigma$ and $2-\sigma$ likelihood contours for the model parameters using Pantheon data sets}\label{f2}
\end{figure}

The obtained best fit ranges of the model parameters are $n=1.1472^{+0.0028}_{-0.00042}$ and $\beta=-8862.2 \pm 1.0 $.

\subsubsection{H(z)+Pantheon datasets}

The  $\chi^{2}$ function for the H(z)+Pantheon data sets is given as 

\begin{equation}
\chi^{2}_{total}= \chi^{2}_H + \chi^{2}_{SN}
\end{equation}

The $1-\sigma$ and $2-\sigma$ likelihood contours for the model parameters $n$ and $\beta$ using H(z)+Pantheon data set is presented below.

\begin{figure}[H]
\includegraphics[scale=0.85]{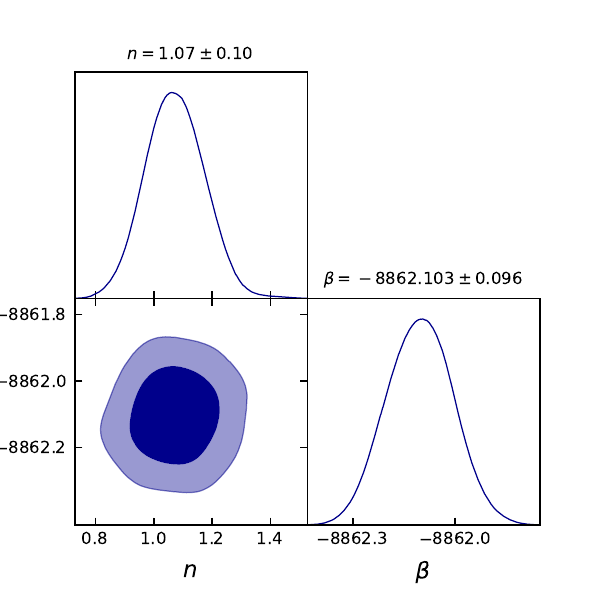}
\caption{The $1-\sigma$ and $2-\sigma$ likelihood contours for the model parameters using H(z)+Pantheon data sets}\label{fa}
\end{figure}

The obtained best fit ranges of the model parameters are $n=1.07 \pm 0.10$ and $\beta=-8862.103 \pm 0.096 $.

The evolution profile of the density and deceleration parameter corresponding to the constrained values of the model parameters are presented below.

\begin{figure}[H]
\includegraphics[scale=0.49]{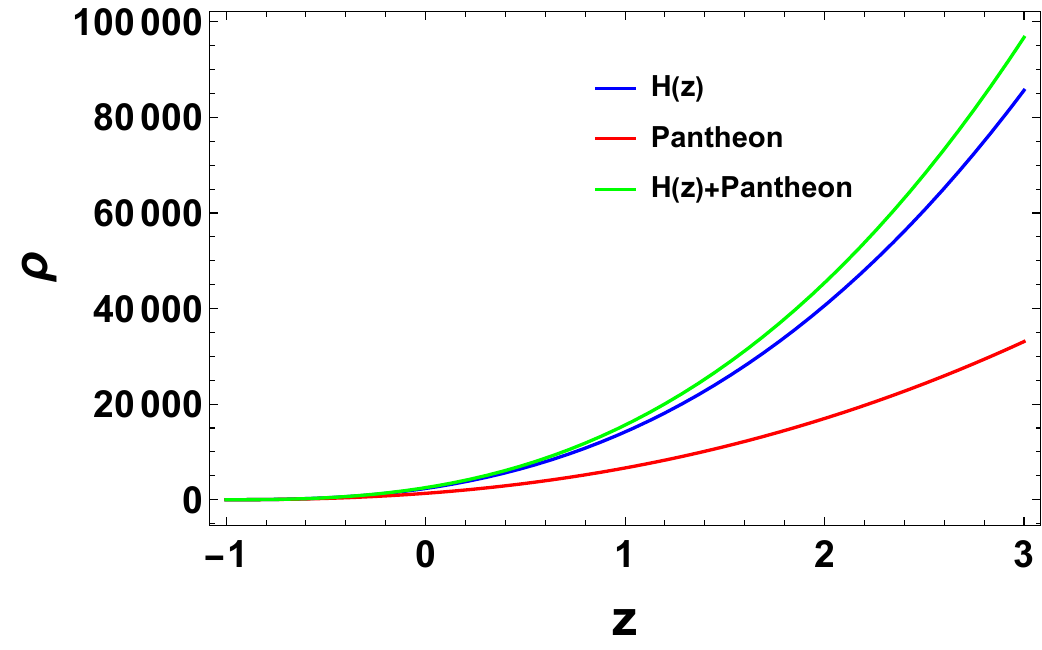}
\caption{Profile of the density vs redshift.}\label{f3}
\end{figure}

\begin{figure}[H]
\includegraphics[scale=0.49]{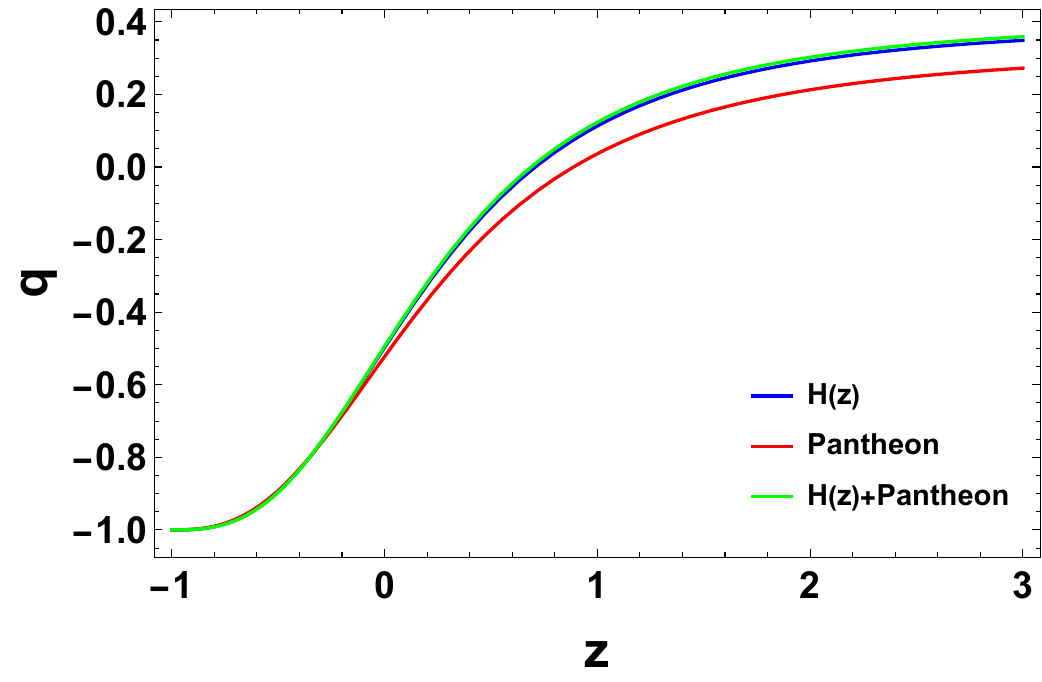}
\caption{Profile of the deceleration parameter vs redshift.}\label{f4}
\end{figure}

Fig. \ref{f3} indicates that the energy density of the cosmic fluid shows positive behavior and it vanishes in the far future. Further, the evolution profile of deceleration parameter in Fig. \ref{f4} reveals that our universe has been experienced a transition from decelerated phase to accelerated phase in the recent past. The transition redshift corresponding to the values of the model parameters constrained by $H(z)$, Pantheon, and the combine H(z)+Pantheon data sets are $z_t=0.708^{+0.029}_{-0.031}$, $z_t=0.887^{+0.0075}_{-0.0009}$, and $z_t=0.688^{+0.262}_{-0.224}$ respectively. Moreover, the present value of the deceleration parameter are $q_0=-0.497^{+0.005}_{-0.004}$ for the $H(z)$ data sets, $q_0=-0.5223^{+0.00003}_{-0.0008}$ for the Pantheon data sets, and $q_0=-0.494^{+0.05}_{-0.035}$ for the H(z)+Pantheon data sets.

\section{Perturbation Analysis of Hubble Parameter}\label{sec6}

In this section, we are going to investigate stability of obtained solution of our proposed $f(R,L_m)$ model under the observational constraint. We have considered a linear perturbation of the Hubble parameter $H(z)$ as

\begin{equation}\label{30}
H^\ast(z)= H(z)(1+\delta(z))
\end{equation}

Here $H^\ast(z)$ represents perturbed Hubble parameter and $\delta(z)$ represents the perturbation term. 

Now by using equation \eqref{21} and \eqref{30} in the matter conservation equation \eqref{c}, we obtained the following expression

\begin{widetext}
\begin{multline}\label{31}
\left(\frac{(z+1)^{\frac{3 n}{2 n-1}} \left(\beta +H_0^2 (6 \delta (z)+3)-2 \beta  \left((z+1)^{\frac{3 n}{1-2 n}}-1\right) \delta (z)\right)}{2 n-1}\right)^{1/n} \\ 
\times \left(\frac{(z+1) \left(2 \delta '(z) \left(3 H_0^2 (z+1)^{\frac{3 n}{2 n-1}}+\beta  \left((z+1)^{\frac{3 n}{2 n-1}}-1\right)\right)+\frac{3 n \left(\beta +3 H_0^2\right) (z+1)^{\frac{n+1}{2 n-1}} (2 \delta (z)+1)}{2 n-1}\right)}{\beta +(2 \delta (z)+1) \left(3 H_0^2 (z+1)^{\frac{3 n}{2 n-1}}+\beta  \left((z+1)^{\frac{3 n}{2 n-1}}-1\right)\right)}-3\right) =0
\end{multline}
\end{widetext}

We solve the equation \eqref{31} numerically since it is highly non-linear and we present the behavior of the perturbation term $\delta(z)$ corresponding to the values of model parameters constrained by observational data sets. 

\begin{figure}[H]
\includegraphics[scale=0.6]{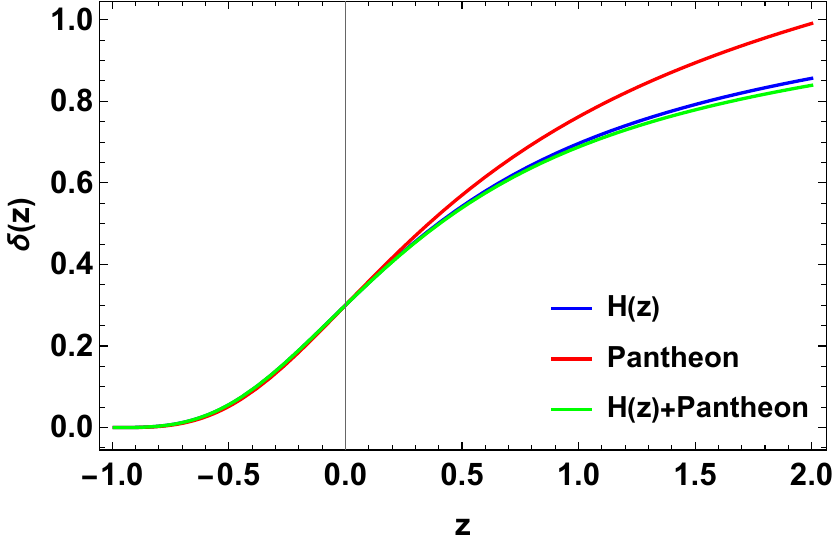}
\caption{Profile of perturbation term $\delta(z)$ corresponding to the values of model parameters constrained by $H(z)$, Pantheon, and the combine H(z)+Pantheon data sets.}\label{f5}
\end{figure}

From \ref{f5} it is clear that, for the constrained values of the model parameters perturbation term $\delta(z)$ decay rapidly at late times. Therefore, the solution of our cosmological $f(R,L_m)$ model shows stable behavior.

\section{Om Diagnostics}\label{sec7}

The Om diagnostic is an effective tool to classify the different cosmological models of dark energy \cite{Om}. It is simplest diagnostic since it uses only first order derivative of cosmic scale factor. For spatially falt universe, it is given as

\begin{equation}
Om(z)= \frac{\big(\frac{H(z)}{H_0}\big)^2-1}{(1+z)^3-1}
\end{equation}

Here $H_0$ is the present value of Hubble parameter. The negative slope of $Om(z)$ correspond to quintessence type behavior while positive slope corresponds to phantom behavior. The constant nature of $Om(z)$ represents the $\Lambda$CDM model. 

\begin{figure}[H]
\includegraphics[scale=0.5]{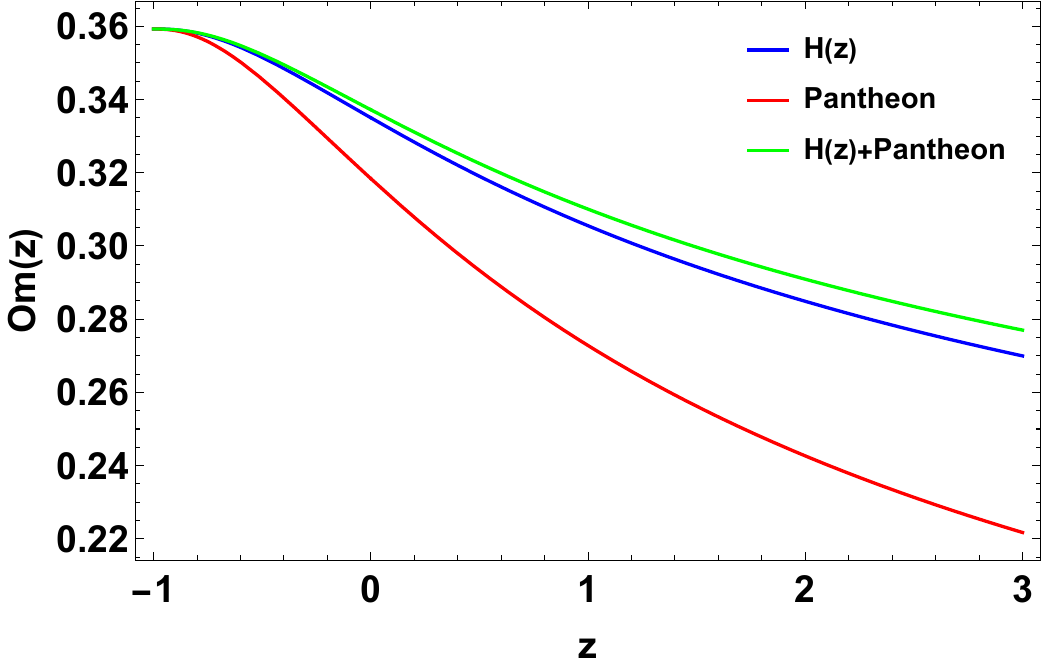}
\caption{Profile of Om diagnostic parameter corresponding to the values of model parameters constrained by $H(z)$, Pantheon, and the combine H(z)+Pantheon data sets.}\label{f6}
\end{figure}

From Fig \ref{f6} we observed that the Om diagnostic parameter for the set of constrained values of the model parameters have negative slope on the entire domain. Thus from Om diagnostic test we can conclude that our cosmological $f(R,L_m)$ model represents quintessence type behavior .

\section{Conclusion}\label{sec8}

In this work, we investigated the late time cosmic expansion of the universe in the framework of $f(R,L_m)$ gravity theory. We considered a non-linear $f(R,L_m)$ model, specifically, $f(R,L_m)=\frac{R}{2}+L_m^n + \beta$, where $n$ and $\beta$ are free model parameters. Then we derived the motion equations for flat FLRW universe. We found the analytical solution presented in equation \eqref{21} for our cosmological $f(R,L_m)$ model. Further, we obtained the best fit values of the model parameters by using $H(z)$ data sets and recently published Pantheon data sets along with the combine H(z)+Pantheon data sets. The obtained best fit values are $n=1.078^{+0.012}_{-0.013}$ and $\beta=-8862.13 \pm 0.99$ for the $H(z)$ datasets, $n=1.1472^{+0.0028}_{-0.00042}$ and $\beta=-8862.2 \pm 1.0 $ for the Pantheon datasets, and $n=1.07 \pm 0.10$ and $\beta=-8862.103 \pm 0.096 $ for the H(z)+Pantheon datasets. In addition, we have investigated the behavior of energy density and deceleration parameter for the constrained values of model parameters. The evolution profile of the deceleration parameter in Fig \ref{f4} indicates a recent transition of the universe from decelerated to accelerated phase and the energy density in Fig \ref{f3} show positive behavior, which is expected. The transition redshift corresponding to the values of the model parameters constrained by $H(z)$, Pantheon, and the combine H(z)+Pantheon data sets are $z_t=0.708^{+0.029}_{-0.031}$, $z_t=0.887^{+0.0075}_{-0.0009}$, and $z_t=0.688^{+0.262}_{-0.224}$ respectively. Moreover, the present value of the deceleration parameter are $q_0=-0.497^{+0.005}_{-0.004}$ for the $H(z)$ data sets, $q_0=-0.5223^{+0.00003}_{-0.0008}$ for the Pantheon data sets, and $q_0=-0.494^{+0.05}_{-0.035}$ for the H(z)+Pantheon data sets. Furthermore, we investigated the stability of the obtained solution of our model under the observational constraint by considering a linear perturbation of the Hubble parameter. From Fig \ref{f5}, we conclude that for the set of constrained values of the model parameters, the obtained solution of our cosmological $f(R,L_m)$ model shows stable behavior. Finally, the evolution profile of the Om diagnostic parameter presented in Fig \ref{f6} indicates that our cosmological $f(R,L_m)$ model follows quintessence scenario . We also found that our cosmological $f(R,L_m)$ model agrees with the constraint $\frac{f_{Lm}(R,L_m)}{f_R(R,L_m)} > 0$ derived in \cite{WG}, which is nothing but $n > 0$ for our considered model.

\section*{Data Availability Statement}

There are no new data associated with this article.

\section*{Acknowledgments} \label{sec10}
L.V.J. acknowledges University Grant Commission (UGC), Govt. of India, New Delhi, for awarding JRF (NTA Ref. No.: 191620024300). R.S. acknowledges UGC, New Delhi, India for providing Junior Research Fellowship (UGC-Ref. No.: 191620096030). SM acknowledges Department of Science and Technology (DST), Govt. of India, New Delhi, for awarding Senior Research Fellowship (File No. DST/INSPIRE Fellowship/2018/IF18D676.)


\end{document}